# An approach to 1000 T using the Electro-Magnetic Flux Compression

**Daisuke Nakamura\*, Hironobu Sawabe, and Shojiro Takeyama**

Institute for Solid State Physics, University of Tokyo, 5-1-5, Kashiwanoha, Kashiwa, Chiba 277-8581, Japan
\*Correspondence and requests for materials should be addressed to D.N. (email: dnakamura@issp.u-tokyo.ac.jp)

**ABSTRACT**

An ultra-high magnetic field was generated by the electro-magnetic flux compression technique under a reduced seed magnetic field condition and achieved maximum magnetic field intensity was investigated. An ordinal pickup coil measurement fails due to the dielectric breakdown at around 500 T. On the other hand, by utilizing the magneto-optical Faraday rotation method with a small probe, the measureable maximum magnetic field increased significantly. It was found that reduced seed field increases the maximum magnetic field, but with a reduced size of the final bore. A highest magnetic field over 763 T and possibly up to 985 T approaching 1000 T was detected.

**Introduction**

An ultra-high magnetic field above 100 T can be artificially generated in the pulsed mode by using techniques such as the single-turn coil (STC)[1], electro-magnetic flux compression (EMFC)[2,3], chemical-explosive flux compression (CE-FC)[4-6], laser-driven flux compression (LD-FC)[7-9], laser-driven single-turn coil target (LD-STC)[10-12], and focused plasma (FP)[13-15]. An extremely short pulse width of the magnetic field and a small bore-size of the magnet in the LD-FC, LD-STC, and FP techniques make it difficult to perform solid-state physics experiments, in which a pulse width of several microseconds and a final bore size of several millimetres are inevitable. These conditions are fulfilled for the STC, EMFC, and CE-FC techniques. However, the CE-FC technique causes a destructive explosion of the magnet and the surrounding set-ups, and from the perspective of safety, refrains from indoor applications. On the other hand, the STC and EMFC techniques, by using a fast high-energy condenser bank system as the energy source possess high controllability and are suitable for indoor applications.

To achieve magnetic fields more than 300 T for solid-state physics experiments, the EMFC technique, which was described in detail by Miura and Herlach[16], is the only method that is currently available. The electrical current flowing in the primary coil compresses the liner set inside the primary coil, and the seed magnetic flux in the liner is simultaneously compressed. A copper-lined primary coil (CL coil) was developed in 2011, and a maximum magnetic field of 730 T (which is a world record) was achieved using a seed magnetic field ($B_s$) of 3.8 T[3]. The homogeneous spatial distribution of magnetic fields (3% in 2-mm lengths at the centre of the coil) is maintained up to 500 T[17]. Therefore, the EMFC technique has so far served as a platform for studying solid-state physics under extreme physical conditions of ultra-high magnetic fields[18-22].

Recently, a precise calibration of the magnetic field intensity generated by the EMFC technique was conducted using the magneto-optical (Faraday rotation (FR)) method[23]. We have shown that the FR angle ($\theta_{FR}$) of fused quartz works well as a reliable probe for measuring magnetic fields of up to 700 T or more, because the FR method is immune to the electromagnetic noise associated with the pulsed magnetic field. The present study aims to achieve a higher magnetic field using the EMFC instrument by reducing the seed magnetic field intensity, where the conventional pickup coil has failed to measure the magnetic flux on the way of the flux compression process. The FR measurement was performed using a probe having a small diameter to resolve this issue.

## Results

A numerical calculation of the liner implosion process was performed based on the finite element method[17] to investigate the dependence of the kinetics of magnetic flux compression on the intensity of $B_s$. Figure 1(a) shows the calculated magnetic field curve $B(t)$ and the inner diameter of the liner $d_L(t)$ for $B_s$ = 2.4, 3.0, 3.2, and 3.8 T. Here, $d_L(t)$ shows an almost identical dependence on time regardless of the difference in values of $B_s$ until 2 μs ahead of the peak field, $B_{turn}$ (shown by down arrows), and then, with the decrease in $B_s$, $d_L(t)$ attains smaller values much faster. This is because the outward magnetic pressure due to the magnetic flux inside a liner enlarges the final inner diameter of the imploding liner ($d_{min}$). Therefore, the magnetic flux inside the liner is compressed more efficiently to reach higher values of $B_{turn}$ in the case of smaller values of $B_s$.

The above calculation was evidenced by the results of the pickup coil experiments #1–#3, as shown below. Figure 1(b) shows the $B(t)$ of the pickup coils and the $d_L(t)$ evaluated from the liner's framing photo images. The grey-colored data in $B(t)$ are those induced by a noise after the destruction of the pickup coil. The velocity of the liner implosion is almost the same (~2.7 km/s) up to 39 μs, whereas there is considerable difference in $B(t)$ around its maximum with a difference in $B_s$. For $B_s$ = 3.8 T (#1, dotted curve), a tiny but clear turn-around peak structure is seen in $B(t)$. On the other hand, in the cases of $B_s$ = 3.2 T (#2, thin solid curve) and 2.4 T (#3, thick solid curve), no turn-around peaks were observed in the pickup coil signals at $B_{cut} \sim 500$–550 T, suggesting that the pickup coils were damaged before reaching $B_{turn}$. This is caused by the dielectric breakdown of the pickup coil due to a huge shockwave and an electromagnetic noise from an imploding liner. Such interference becomes strong with the decrease in $d_L$, when $B_s$ decreases.

Figure 2(a) shows the result of the FR experiment #4 ($B_s = 3.8$ T). The s- and p-polarized components of the FR signal ($V_s$, $V_p$) and the summation of the transmission signals ($V_s + V_p$, multiplied by 0.5 for clarity) are shown. After 40 μs, the transmission signal showed an abrupt decrease, which is caused by the interference of the imploding liner with the quartz rod, and the optical path is interrupted. Therefore, the FR signal was reconstructed by normalizations of $V_s / (V_s + V_p)$ and $V_p / (V_s + V_p)$, as shown in Fig. 2(b). Then, the data were converted into magnetic field intensities $B_{FR} = \theta_{FR}/ VL$, which are shown in Fig. 2(c). The small step-like jump noticed at 38.6 μs in $B_{FR}$ is an artefact caused by an enhanced error in the calculation of $\theta_{FR}$ at $V_sV_p \sim 0$ (the dotted curve is a guideline). The intensity of the magnetic field obtained from the pickup coil ($B_p$) becomes lower than that from the FR angle of the quartz rod at the high magnetic field region. This is already known to be caused by a high-frequency electrical loss in the long electrical coaxial cable used for signal transmission[23].

FR experiment #5 (with reduced $B_s$) was attempted to obtain an increase in $B_{FR}$ with a small $d_L$. The outer diameter of the quartz rod holder was reduced to 2.0 mm (Supplementary Fig. S1). The s- and p-polarized FR signals ($V_s$, $V_p$) of experiment #5 ($B_s = 3.0$ T) are shown in Fig. 3(a) (the abscissa is expanded in the inset). Although there is a considerably weak optical transmission, an oscillation of the FR signal is seen in $V_p$. The extremal points of the oscillation are indicated by the circled number n, where $\theta_{FR}(t)$ is described by 90*n-45 deg. $V_s$, which is initially close to zero level, is used as a check of the background contribution of light coming from the imploding high-temperature liner. It is known that when the imploding liner approaches $B_{turn}$, intense illumination flashes from the inner surface of the liner, as seen in the snapshot of the framing camera (Supplementary Fig. S2, taken at 39.8 and 40.8 μs). This illumination appears as a time-evolving background signal after 39.5 μs in Fig. 3(a).

On going from n = 10 to n = 11, the intensity of $V_p$ decreases, whereas the background intensity in $V_s$ increases monotonically. Therefore, the extremum at n = 11 is regarded as that of an intrinsic FR oscillation, since $V_p$ at n = 11 is almost at zero level. Considering the initial $\theta_{FR}$ setting to be 18.7°, the magnetic field intensity at n = 11 is evaluated as (90*11-45-18.7)/(2.179*0.559)+3.0 = 763 T.

**Discussion**

It is further examined whether the extrema n > 11 in Fig. 3(a) is a real signal or simply a noise. By subtracting the background contribution, the oscillation of $V_p(t)$ becomes more obvious in Fig. 3(b). Up to n = 13, the oscillation period of $V_p$ decreases gradually, which turns out the sharp rise of the B(t) curve. Then, between n = 13 and n = 14, the oscillation period becomes broad, suggesting the slowing-down of the $B(t)$ curve. In general, the waveform of the magnetic field determined from the FR oscillation behaves quite similar to the waveform of the magnetic field generated by the flux-compression process[3]. As a precursor of the turn-around phenomenon in the EMFC technique, the $B(t)$ curve shows typical "slow-down phenomenon" just before the turn-around peak. Hence, the 12–14$^{th}$ extrema of $V_p$ can be regarded as an intrinsic FR oscillation.

The magnetic field intensities evaluated from the extremal points of $V_p$ in FR experiment #5 are plotted up to the $n^{th}$ extrema in Figs. 4(a)–(d) (open symbols). The $B(t)$ curve of #2 ($B_s$ = 3.2 T, dotted marks in Figs. 4(a)–(d)) almost reproduces the $B(t)$ curve of #5 up to 550 T, which supports the validity of the data analysis. In Fig. 4(a), the $B(t)$ curve of experiment #5 does not show such distinct "slow-down phenomenon" prior to the turn-around phenomenon. This suggests that $B_{turn}$ should be considerably higher than 763 T. Therefore, the maximum magnetic field intensity may reach the value at the extremal points $n = 12$ (837 T), $n = 13$ (911 T), or even $n = 14$ (985 T), as demonstrated in Figs. 4(b)–(d). It is to be noted that the "slow-down phenomenon" appears in Fig. 4(d). Figure 4(d) presents the most probable magnetic field curve in experiment #5, achieving a maximum magnetic field above 985 T, very close to 1000 T.

The dependence of the compressed magnetic field intensity on $B_s$ is demonstrated in Fig. 5. Figure 5(a) presents the maximum magnetic field intensity measured by the pickup coil. The closed circles indicate the data obtained in the present study (#2, #3). The open squares indicate the data taken from Ref. 3, in which there is an optimal value of $B_s$ (3.6–4.1 T) for obtaining the highest magnetic field intensity using the EMFC instrument (dashed curve is a guide). At $B_s$ = 2.4–3.5 T, the pickup coil breaks at $B_{cut}$ (for a distinction between $B_{cut}$ and $B_{turn}$ against data in Ref. 3, see Supplementary Fig. S3). Therefore, there is a threshold value of $B_s$ around 3.5–3.8 T, below which it is difficult to measure $B_{turn}$ successfully using the present pickup coil technique, due to the extremely small size of $d_{min}$.

On the other hand, as shown in Fig. 5(b), the result of the FR measurements indicates that $B_{turn}$ ($B_s$ = 3.0 T) might be significantly higher than $B_{turn}$ ($B_s$ = 3.8 T), which is qualitatively consistent with the numerical result shown as a dashed-dotted curve. For experiment #5, the possible candidates of the maximum magnetic field are presented as closed diamonds, which are evaluated in Figs. 4(a)–(d). As the numerical calculation is underestimated compared with $B_{turn}$ at $B_s \geq 3.8$ T, it is quite plausible that $B_{turn}$ at $B_s$ = 3.0 T exceeds 985 T, and is nearly 1000 T.

In conclusion, we examined the maximum magnetic field intensity obtained by the EMFC technique with regard to a reduced seed magnetic field. We found that the EMFC technique could provide higher maximum magnetic fields when the seed magnetic field is reduced, but with smaller final bore. The conventional pickup coil fails to detect the final maximum field due to its destruction by the imploding liner at the very end of the flux compression process. By performing FR measurement of a quartz rod having a small diameter in the EMFC experiment with the seed field reduced to 3 T, the maximum magnetic field detected is significantly increased at least up to 763 T, and most probably up to more than 985 T, close to 1000 T.

**Methods**

The EMFC technique was used to generate the ultra-high magnetic fields. The experimental conditions are listed in Table 1. The energy of 4.0 MJ stored in the condenser bank is injected into the primary CL coil. In this study, we performed the EMFC experiments for $B_s$ = 3.2 T (#2), 2.4 T (#3), and 3.0 T (#5), respectively. For comparison, the data of EMFC experiments for $B_s$ = 3.8 T were taken from Ref. 3 (#1) and Ref. 23 (#4). The other conditions, including the primary CL coil dimensions, were the same as in #1 (experiment "a" in Ref.3) and in #4 (experiment "fast"-type in Ref. 23).

In experiments #1–#3, the liner inner diameter ($d_L$) was estimated from the liner motion taken by a high-speed framing camera (Imacon 468, John Hadland ctd.) with a time resolution of 10 ns. The speed of the imploding liner was compared. For the magnetic field probe, a pickup coil with a diameter of 1 mm was wound around the G10 rod, which was set at the centre of the liner.

In experiments #4 and #5, a fused quartz rod was prepared as the magnetic field probe for the FR measurements. The diameters of the quartz rods were 2.0 mm (#4) and 1.1 mm (#5), and the lengths $L$ were 0.618 mm (#4) and 2.179 mm (#5), respectively. The schematic diagram of the FR measurement is provided in Ref. 23 (also see Supplementary Fig. S4). The semiconductor laser (coherent "CUBE") was used as a light source (#4: 638 nm, #5: 404 nm). The magnetic field intensity was calculated from the FR angle using the relation $B = \theta_{FR}/VL$, where the Verdet constant ($V$) of the fused quartz is 0.200 deg./mmT (638 nm) and 0.559 deg./mmT (404 nm)[23] (also see Supplementary Fig. S5). By comparing the $B(t)$ of a pickup coil with that of an FR probe, $V$ was evaluated with an accuracy of 3%, which is mostly limited by the estimation error of the pickup coil cross-sectional area (i.e., diameter)[23].

The datasets generated during and/or analysed during the current study are available from the corresponding author on reasonable request.

**Acknowledgements**

**Author contributions**
D.N. conducted the experiments and analysed the data. H.S. provided support in the experiments. D.N. wrote the manuscript. S.T. is a coordinator of the project and reviewed the manuscript.

**Competing interests**
The authors declare no competing financial interests.

**Figure legends**

**Figure 1.** (a) Numerical calculation results of the magnetic fields (curves) and the motion of the inner wall of the liner (markers, $d_L$) for the seed magnetic field $B_s$ = 2.4, 3.0, 3.2, and 3.8 T, respectively. (b) Results of EMFC experiment. Curves show $B(t)$ (#1: dotted, #2: thin solid, and #3: thick solid) and the marker shows the inner diameter of the liner, $d_L$ (#1: open circle, #2: closed square and #3: closed circle). The dashed line guides the speed of the liner as 2.7 km/s.

**Figure 2.** Results of the FR experiment #4 ($B_s$ = 3.8 T). (a) A transient record of the optical transmission intensities of $V_s$ and $V_p$. The dotted line indicates the values of $(V_s + V_p)/2$. (b) $V_{s,n}$ and $V_{p,n}$, obtained as a result of normalization and (c) the magnetic field curves obtained by the FR measurement ($B_{FR}(t)$) and that of a pickup coil ($B_p(t)$).

**Figure 3.** Results of the FR experiment #5 ($B_s$ = 3.0 T). (a) s- and p-polarized light intensities of FR measurement, $V_p$ and $V_s$. The inset displays the figure with its abscissa expanded. The circled numbers indicate the extremal points of $V_p$ as an oscillation of the FR signal. (b) Replot of the FR signal after subtraction of the background signal in (a). For comparison with $V_s$, the signal of $V_p$ is multiplied by 0.2.

**Figure 4.** Magnetic field curves of experiment #5 ($B_s$ = 3.0 T, open symbols) and of experiment #2 ($B_s$ = 3.2 T, dotted curve). The extremal points with circled number 11–14 of $V_p$ in Fig. 3 are assumed to present $B_{max}$ equal to (a) 763 T, (b) 837 T, (c) 911 T, and (d) 985 T, respectively.

**Figure 5.** Seed magnetic field dependence of the maximum magnetic field intensity. (a) The result of the pickup coil measurements. The closed circles indicate the data of experiments #2 and #3. The open squares indicate data taken from Ref. 3, and the dashed curve acts as a guide for the eye. (b) The result of FR measurements. For experiment #5, the candidates with the maximum magnetic field intensity are plotted with circled numbers from 10 to 14 (closed diamonds). The dashed-dotted curve displays the result of our simulation.

**Tables**

| Experiment | #1[3] | #2 | #3 | #4[23] | #5 |
|---|---|---|---|---|---|
| Seed field $B_s$ | 3.8 T | 3.2 T | 2.4 T | 3.8 T | 3.0 T |
| Method | Pickup coil & framing camera | | | FR & pickup coil | FR |
| Diameter of quartz | - | | | 2.0 mm | 1.1 mm |

**Table 1**. Summary of the present EMFC experimental conditions.

**Figures**

**Fig. 1**

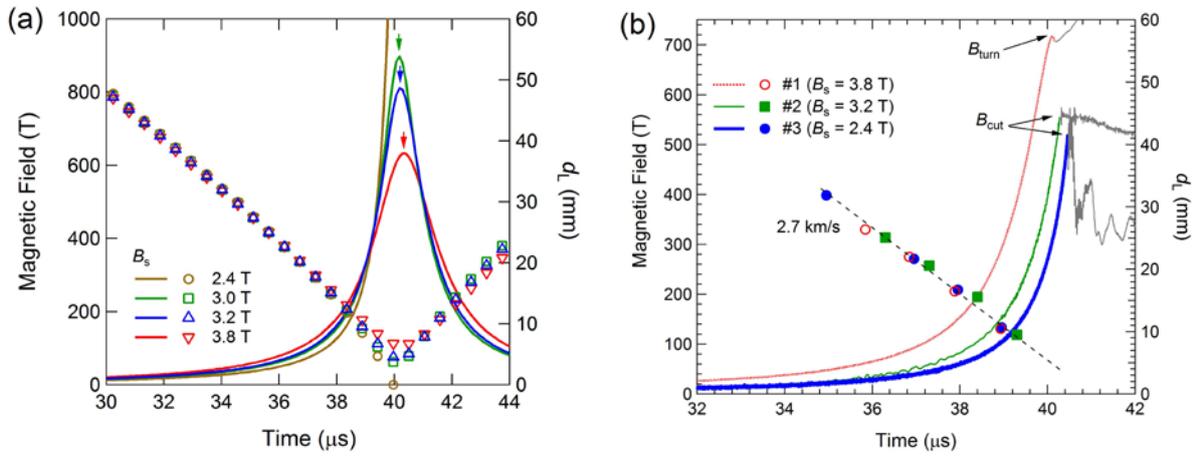

**Fig. 2**

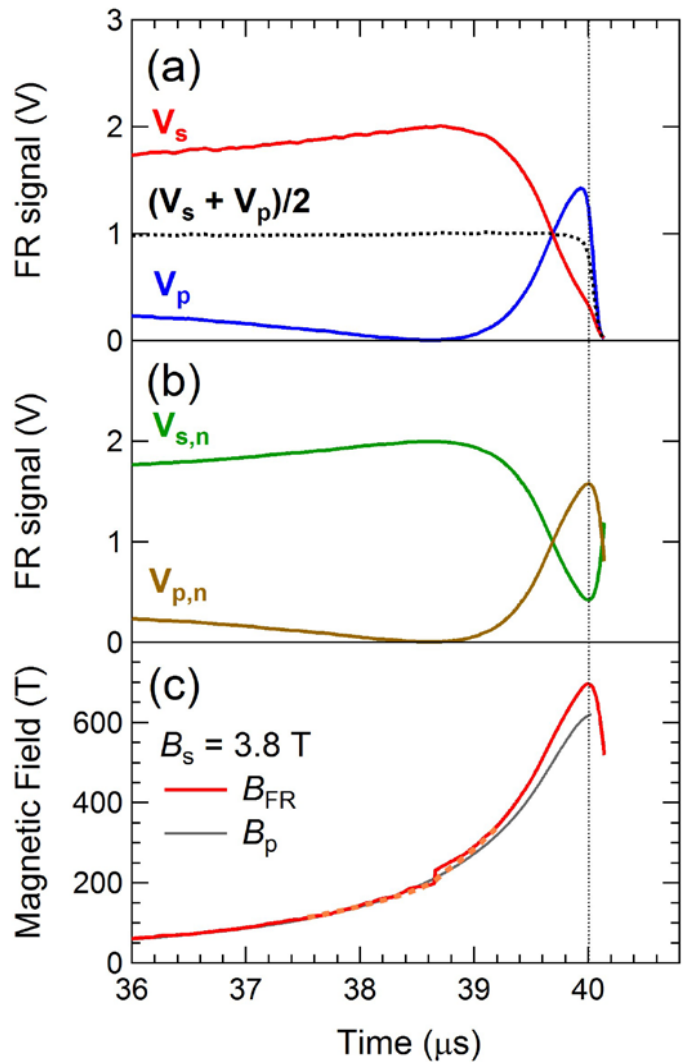

**Fig. 3**

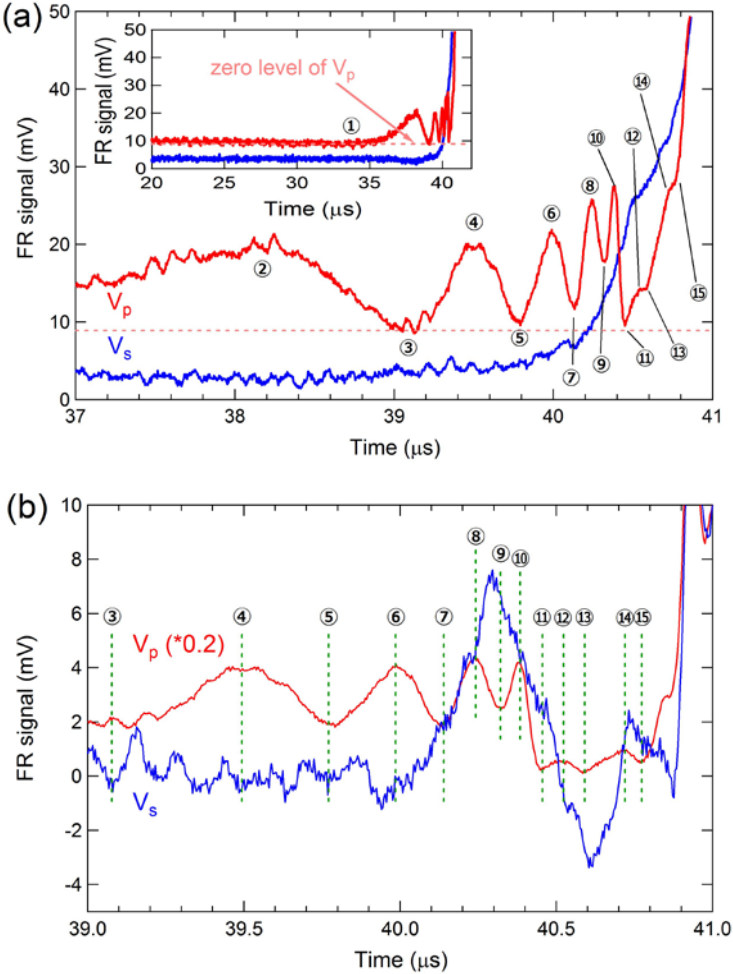

**Fig. 4**

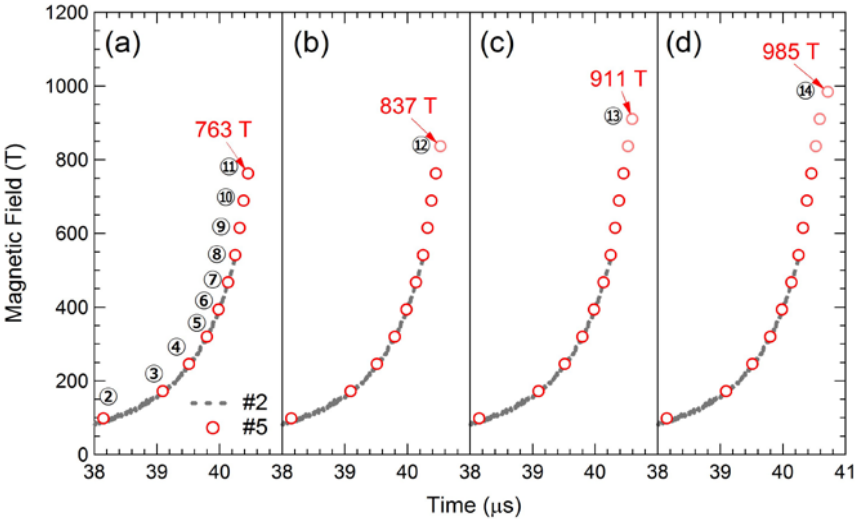

**Fig. 5**

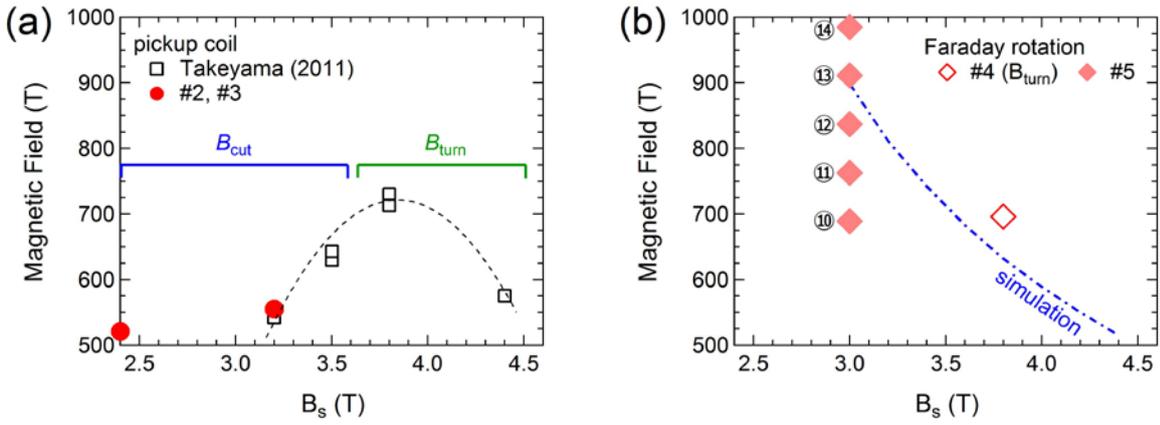

**Supplementary information for**

**An approach to 1000 T using the Electro-Magnetic Flux Compression**

**Daisuke Nakamura*, Hironobu Sawabe, and Shojiro Takeyama**

Institute for Solid State Physics, University of Tokyo, 5-1-5, Kashiwanoha, Kashiwa, Chiba 277-8581, Japan

*E-mail: dnakamura@issp.u-tokyo.ac.jp

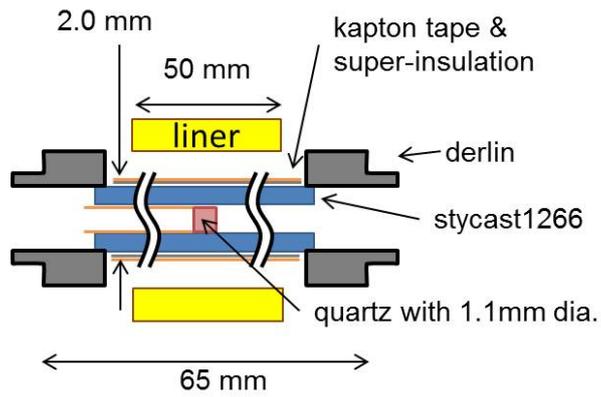

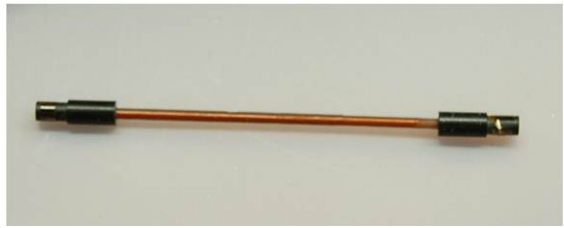

**Figure S1.** (top) Schematics and (bottom) photograph of a quartz rod holder used in FR measurement #5.

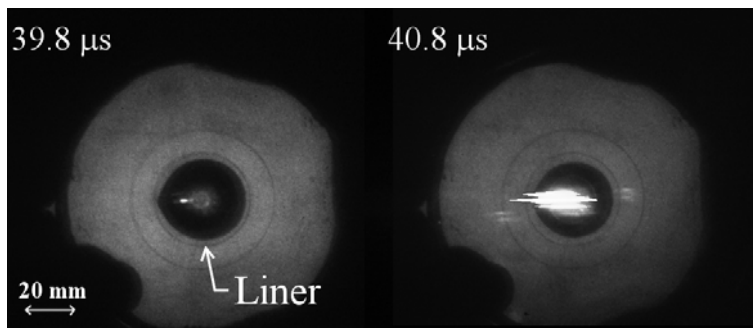

**Figure S2.** Framing photos of the liner taken at almost the end of the implosion before the peak field.

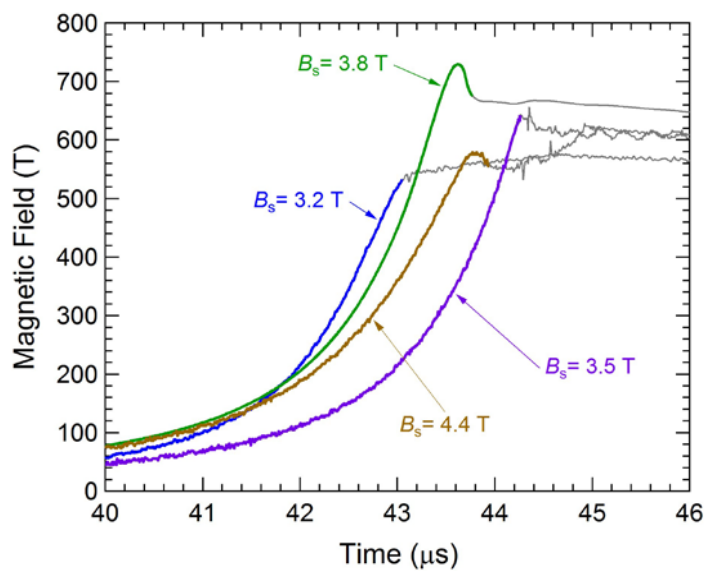

**Figure S3.** Magnetic field curves of the EMFC experiments in Fig. 7 of Ref. 3. The thin grey region in $B(t)$ is the extrinsic noise, after the pickup coil is broken.

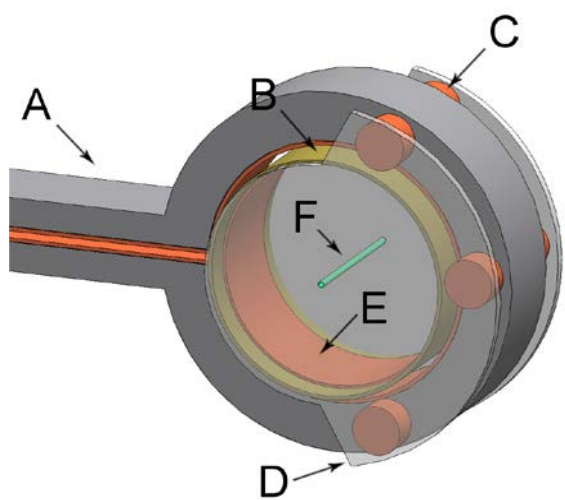

**Figure S4.** Schematic illustration of the primary coil used in the electro-magnetic flux compression technique. A: copper-lined primary coil. B: vacuum chamber. C: spacer. D: vacuum flange. E: liner. F: quartz rod holder.

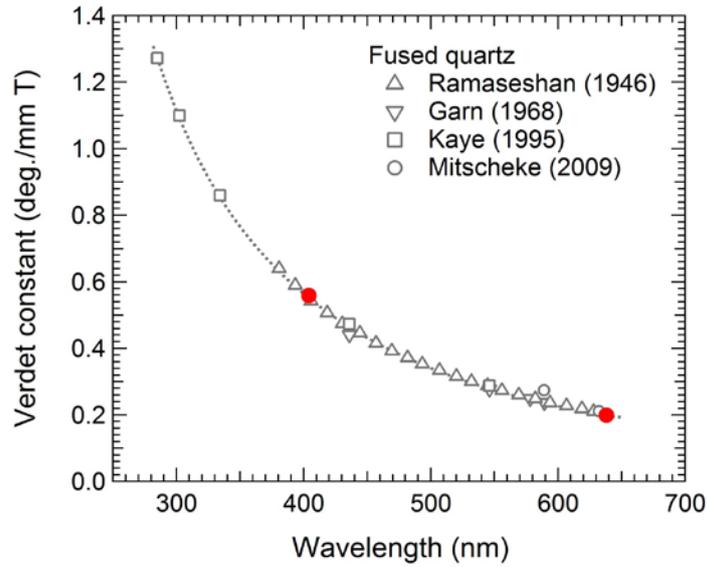

**Figure S5.** The wavelength dependence of the Verdet constant of fused quartz. The closed symbols are the Verdet constant used in this study. The dotted curve is a guideline.

**Supplementary reference**

1. Ramaseshan, S., Proc. Indian Acad. Sci., Sec. A **24**, 426 (1946).
2. Garn, W. B., Caird, R. S., Fowler, C. M., & Thomson, D. B., Rev. Sci. Instrum. **39**, 1313 (1968).
3. Kaye, G. W. C. & Laby, T. H. (eds.). *Tables of Physical and Chemical Constants and Some Mathematical Functions* (16th edition) 143 (Longmans, London, 1995).
4. Mitscheke, F. *Fiber Optics: Physics and Technology* (Springer, 2009).